\documentclass[pra,twocolumn,showpacs,preprintnumbers,amsmath,amssymb,revtex4-1]{revtex4-1}
\usepackage{amsfonts, amsmath, bm, bbm, graphicx, epsfig, epstopdf}
\usepackage{dcolumn}
\usepackage{textcomp}
 \usepackage[caption=false]{subfig}
 \usepackage[caption=false]{subfig}

    \usepackage{braket}
\renewcommand\bra[1]{{\langle{#1}|}}
\makeatletter
\renewcommand\ket[1]{%
  \@ifnextchar\bra{\k@t{#1}\!}{\k@t{#1}}%
}
\newcommand\k@t[1]{{|{#1}\rangle}}
\makeatother

\usepackage{subfig}

\begin{document}

\bibliographystyle{apsrev}

\title{ Fast multi-component cat generation under resonant or strong dressing Rydberg Kerr interaction}

\author{Mohammadsadegh Khazali}
\affiliation{Department of Physics, University of Tehran, Tehran, 14395-547, Iran}
\affiliation{mskhazali@ut.ac.ir}

\begin{abstract}
Cat states are maximally entangled states with applications in metrology and fault-tolerant quantum computation.
The experiments have revealed that Rydberg collective avalanche decoherence acts as the bottleneck for cat creation with Rydberg atoms. This process initiates after the black body radiation (BBR-)induced decay of Rydberg atoms and sets a strong limit on the cat creation time. These findings necessitate the exploration of new ideas to accelerate current Rydberg cat schemes.
To enhance the interaction-to-loss ratio, this paper delves into cat state formation in the strong Rydberg dressing regime, uncovering the emergence of cat states despite the presence of complex orders of nonlinearities. This unexplored regime demonstrates the potential for rapid cat state formation, particularly beneficial for operation in typical 2D lattices in Rydberg Labs. In an extreme case, this article demonstrates that second-order nonlinearity could be isolated under resonant Rydberg driving if a large number of atoms are accommodated inside the blockade volume. The resonant model significantly enhances the interaction-to-loss ratio while circumventing the adiabaticity condition, allowing fast switching of lasers. In addition, the paper presents a method for generating multi-component cat states, which are superpositions of $m$ coherent spin states ($\ket{m-\text{CSS}}$). The maximum value of $m$ is determined by the number of atoms within the blockade radius, where $m=\sqrt{N}$. The states with larger $m$ are more robust against the presence of multiple orders of nonlinearity in the strong dressing Hamiltonian and are accessible in a much shorter time compared to traditional 2-component cat states.
\end{abstract}

\maketitle

\section{introduction}

The nonlinear interaction resulting from Rydberg dressing has potential applications in various areas, including spin squeezing \cite{Eck23, Saf21, Hin23,Mul23, Gil14}, the generation of Schr\"odinger cat states \cite{Omr19, Kha16, Kha18, Han16}, as well as in the fields of many-body physics and quantum materials \cite{Zei16, Hon10, Kha21, Hen10, Mob13, Muk15, Pen20}. Until now, efforts to isolate the quadratic order of nonlinearity have been confined to far-off resonant laser driving that weakly dresses the excited states with strongly interacting Rydberg states, a condition known as the weak dressing regime. This quadratic nonlinearity was assumed as a coherent candidate for making cat states \cite{Kha16, Kha18}. 
Cat states are highly fragile, as a single decay can result in the complete destruction of entanglement.
 Experimental endeavors have encountered a significant obstacle in the form of collective avalanche loss \cite{Gol16, Zei16}, necessitating any cat proposal to operate within a specific time window to keep the chance of BBR-induced decay below 20\% \cite{Zei16}. However, the small interaction-to-loss ratio in the weak dressing regime does not facilitate the generation of even small cat states specifically in the typical 2D lattices available in most Rydberg labs.
To address these challenges, this paper delves into the unexplored strong dressing regime to enhance the interaction-to-loss ratio. Our findings indicate that approaching resonant Rydberg driving significantly boosts the operation speed.

Deviating from the conventional weak dressing regime in typical small ensembles ($N \ll 400$), the exploration of stronger dressing effects amplifies many-body interactions and unlocks higher orders of nonlinearities. In the context of employing two distinct Rydberg dressing methodologies for cat state generation \cite{Kha18, Kha16}, previous studies have highlighted the susceptibility of the Lipkin-Meshkov-Glick model to mixed nonlinearities \cite{Kha18}. 
Conversely, our investigation reveals that in the Yurke-Stoller framework \cite{Yurk86, Kha16} the impact of higher-order nonlinear terms could be less detrimental. 
Taking a novel approach to this challenge, this paper showcases the isolation of quadratic nonlinearity in an extreme scenario specifically, during resonant Rydberg driving when a substantial number of atoms ($N \gtrsim 400$) are enclosed within the blockade radius. This breakthrough leads to a notable enhancement of the interaction-to-loss ratio, marking a significant stride towards the creation of large entangled states.

While there are multiple measures for quantumness, an intriguing figure of merit could be the number of superposition states that elements could possess simultaneously. Yurke-Stoller discussed the formation of two-component cat states $\ket{2-\text{CSS}}$ under a sole even term of nonlinearity $U \propto N_e^{2k}$. This state is equivalent to the superposition of two coherent spin states pointing in opposite directions on the Bloch sphere. They also highlighted the formation of four-component cats $\ket{4-\text{CSS}}$ under an isolated odd order $U \propto N_e^{2k+1}$. This article extends this model to create a superposition of $m$ coherent spin states $\ket{m\text{-CSS}}$ with $m \leq \sqrt{N}$, where $N$ is the number of atoms. Our numerical study reveals that as we move towards a strong dressing regime, cat states with larger $m$ experience less fidelity reduction and have a much shorter creation time than two-component cat states. 
The application of $\ket{m\text{-CSS}}$ state in metrology yields a signal that is periodic in the metrological phase divided by $m$, reducing the inversion region of the dynamic range of the signal.
Another potential use of $\ket{m\text{-CSS}}$ is in error correction with cat qubits. Encoding in the superposition of $2k$ CSS with $k>1$ enables correction of error types that would not be feasible with a $\ket{2\text{-CSS}}$ cat \cite{Maz16}.

Despite the benefits of the strong dressing regime, the absence of an analytical model describing the interaction profile at this regime \cite{Saf21} has hindered mean-field studies on this unexplored range of parameters. Appendix I  investigates an analytic form of the interaction profile involving two and three-body interactions, providing insights into the effects of lattice geometry and controlling parameters on enhancing specific orders of nonlinearities. Furthermore, it opens new opportunities to study the dynamics of Rydberg dressed BEC \cite{Hen10, Kha21} under a stronger dressing regime using mean-field theory.

  \begin{figure}
\centering    
        \subfloat{%
    \includegraphics[width=.47\textwidth]{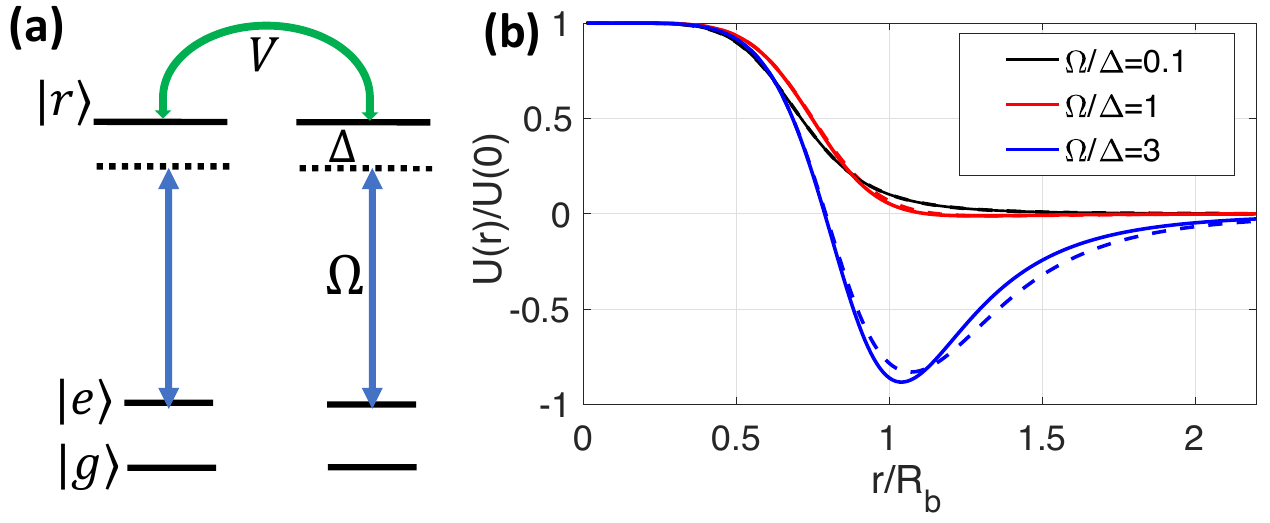}} \hfill           
\caption{Transition from weak to strong Rydberg dressing and resonant driving. (a) Level scheme - the spin of atoms consists of $|g\rangle$ and $|e\rangle$ electronic states. The desired Kerr-type interaction is generated by off-resonant laser driving of atoms in $|e\rangle$ state to the Rydberg level. This paper studies the Kerr-type  interaction in the transition from weak  $\Omega\ll\Delta$  to strong dressing regime $\Omega\gg\Delta$ and  introduces the resonant driving $\Delta=0$ Kerr Hamiltonian.
(b) The binary dressing Interaction profile derived from the steady state of the master equation. At both weak and strong dressing, the homogeneous soft-core is provided for interatomic distances below $R_b/2$. The blockade radius in the weak and strong dressing regimes are defined by $R_b=(C_6/\Delta)^{1/6}$ and $R_b=(C_6/\Omega)^{1/6}$ respectively.  The interaction is scaled by $U_0$ which is the exact dressing interaction for totally blockaded atoms.
}\label{Fig_Scheme}
\end{figure}

The paper is structured as follows: Section \ref{Sec_Dress} elucidates the traditional concept of Rydberg dressing and explores the relation between dressing strength and the order of nonlinear terms within the Hamiltonian. In Section \ref{Sec_mCSS}, the formation of $\ket{m-CSS}$ states is expounded upon.
Furthermore, Section \ref{Sec_Adv} delves into the benefits derived from transitioning from weak to strong dressing techniques. The formulation of the Kerr Hamiltonian for resonant Rydberg driving and its application in cat state generation is detailed in Section \ref{Sec_Resonance}.
Lastly, an analysis of many-body interactions under strong dressing is provided in Appendix I.

\section{Dressing Interaction}
\label{Sec_Dress}

When all atoms are accommodated within the blockade radius, the dressing laser connects the state without Rydberg excitation $|\psi_0\rangle=\otimes_{i}\left|\phi_{i}\right\rangle$ (where $\phi\in{e,g}$) to a state where only one atom in the $|e\rangle$ level is excited to the Rydberg state $|\psi_1\rangle=\sum_{i}\left|\phi_{1}...r_{i}...\phi_{N}\right\rangle$ with a collective Rabi frequency of $\sqrt{N_{e}}\Omega_{r}$ and laser detuning $\Delta$. Here $N_e$ represents the number of atoms in the excited state $\ket{e}$. During the Rydberg dressing process, the collective light-shift experienced by the ground-dressed state could be obtained as 
\begin{equation}
H_{\text{exact}}=\frac{\Delta}{2}(1 - \sqrt{1+\frac{\hat{N}_{e}\Omega_{r}^{2}}{\Delta^2}}).
\label{Eq_exact}
\end{equation}
This light-shift could be Taylor expanded in the weak dressing regime $(\frac{N_e\Omega^2}{\Delta^2}\ll 1)$ as
\begin{equation}
H_{\text{w}}=-\frac{1}{4} \frac{\hat{N}_e \Omega^2}{\Delta} + \frac{1}{16} \frac{\hat{N}_e^2 \Omega^4}{\Delta^3} -\frac{1}{32} \frac{\hat{N}_e^3 \Omega^6}{\Delta^5}+\frac{5}{256} \frac{\hat{N}_e^4 \Omega^8}{\Delta^7}+O(\hat{N}_e^5).
\label{Eq_LightShift}
\end{equation}
Going to stronger dressing with larger $N_e \frac{\Omega^2}{\Delta^2}$ activates the higher orders of non-linearity and raises the many-body interaction terms in the Hamiltonian, see App.~I.

These binary dressing interaction profiles are presented with solid lines in Fig.~\ref{Fig_Scheme}b. It is derived from the steady state of the master equation encountering laser couplings, dipolar Rydberg interaction and spontaneous emission from the Rydberg level, see App.~II.  Dressing interaction features a soft core within the blockade radius $R_b$ over which the interaction makes the two  Rydberg excitations out of resonance with the laser and forms the effective interaction potentials of Eq.~\ref{Eq_exact}.  The blockade radius in the weak and strong dressing regimes are defined by $R_b=(C_6/\Delta)^{1/6}$ and $R_b=(C_6/\Omega)^{1/6}$ respectively.
In the extreme limit of strong dressing where $\Omega>\Delta$,  the interaction would be comparable with the Rabi frequency around $R_b$ and cause blockade leakage. This populates more than one Rydberg atom featuring strong dipolar interaction, represented by an interaction peak in the blue line of Fig.~\ref{Fig_Scheme}b. At further distances the $1/r^6$ van-der Waals tail could be recognized in both weak and strong dressing. Dashed lines in  Fig.~\ref{Fig_Scheme}b represent the interaction of three atoms on an equilateral triangle with sides length $r$.

\section{Making superposition of $m$-CSS}
\label{Sec_mCSS}

A coherent spin state (CSS) is defined as a direct product of single spin states \cite{CSS} 
\begin{equation}
|\theta,\phi \rangle=\otimes _{i=1}^{N}[\cos{\theta}|g \rangle_{i}+\sin{\theta} e^{i \phi} |e \rangle_{i}],
\end{equation} 
where all the spins are pointing in the same direction, and $\phi$ and $\theta$ are the angles on the (collective) Bloch sphere. The CSS can also be represented as \cite{CSS}
\begin{equation}
|\eta \rangle= |\theta,\phi \rangle=(1+|\eta|^{2})^{-N/2} \sum_{N_{e}=0} ^{N} \eta^{N_{e}} \sqrt{C(N,N_e)}|N;N_{e}\rangle,
\label{Eq_ICSS}
\end{equation} 
where $\eta=\tan(\theta/2)e^{-i\phi}$,
$C(N,N_e)\equiv\left(\begin{array}{c} N\\ N_{e} \end{array}\right)$ and  $|N;N_{e} \rangle=\frac{1}{\sqrt{C(N,N_e)}}\sum_{i1<i2<...<iN_{e}}^{N}|g_{1}...e_{i1}...e_{iN_{e}}...g_{N}\rangle$ is the Dicke state of $N_{e}$ excited atoms, where $|N;N_{e} \rangle$ is an alternative representation of the $|J \, M\rangle$ basis with $N=2J$ and $N_e=J+M$.

Considering the time evolution of the CSS of Eq.~\ref{Eq_ICSS} under the dressing Hamiltonian of Eq.~\ref{Eq_LightShift}, the linear term in $N_e$ preserves the CSS and only generates a rotation around $J_z$ while the quadratic term $\chi N_e^2$ causes spin-squeezing over the Bloch sphere and when the CSS was squeezed over the equator it starts to make a superposition of $m$-CSSs at $t_m=\frac{2\pi}{m\chi}$.
\begin{equation}
\label{Eq_mCSS}
\ket{m-\text{CSS}}=\frac{1}{\sqrt{m}}\sum\limits_{k=1}^{m} \text{e}^{i\alpha_k}   \ket{\theta=\pi/2;\phi=k\frac{2\pi}{m}+\phi_0}
\end{equation}
 The values of $\alpha_k$ are obtained numerically in App.~IV. Continuing the interaction at $t_1$, one can observe the revival of the initial CSS. This revival can be used as proof for the successful creation of a quantum superposition at $t_m$ since a statistical mixture of CSSs at $t_m$ would evolve into another mixture of separate peaks, see App.~III and Fig.~\ref{Fig_QFWeakStrong}.

\begin{figure*}
\centering    
        \subfloat{%
    \includegraphics[width=.3\textwidth]{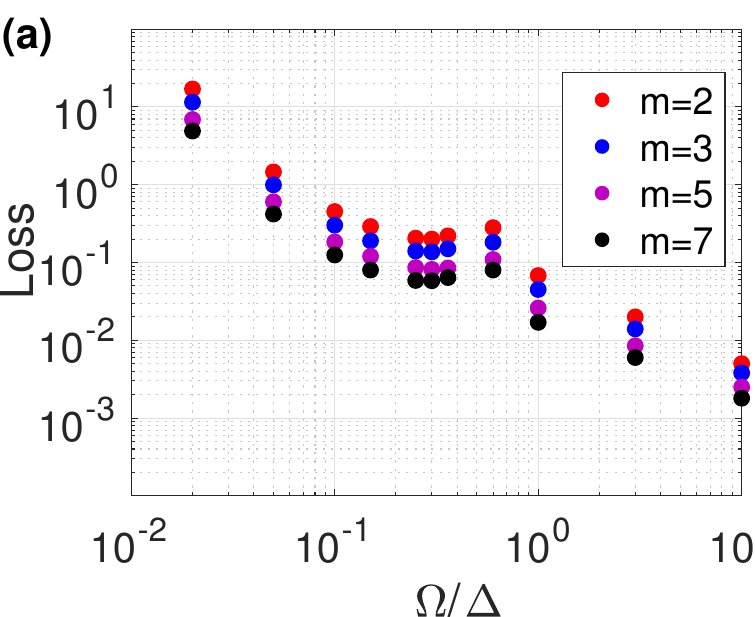}} \hfill  
            \subfloat{%
    \includegraphics[width=.3\textwidth]{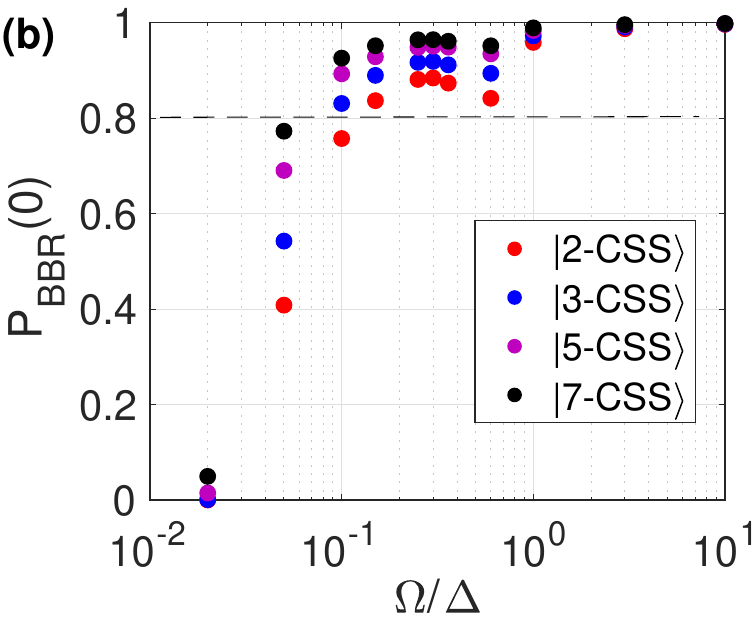}} \hfill      
      \subfloat{%
    \includegraphics[width=.3\textwidth]{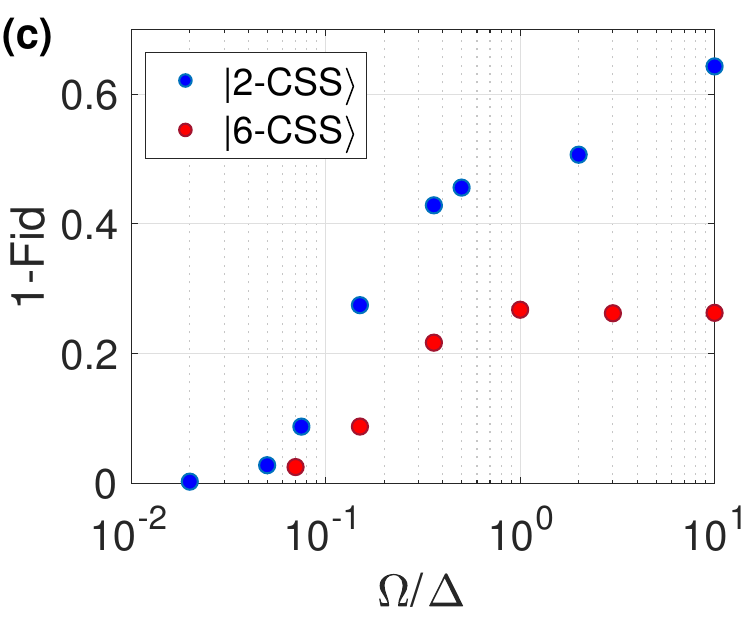}} \hfill 
\caption{Transition from weak to strong dressing. (a)  The population loss  over the $|m-\text{CSS}\rangle$ generation $\text{Loss}=P_r\Gamma_rt_m$ is plotted as a function of dressing strength. Here constant detuning $\Delta/2\pi=0.02$GHz in adressing $|53P_{3/2},3/2\rangle$ state and $N=48$ atom number are considered.  At very weak dressing with quadratic nonlinearity, the interaction to loss ratio scales by $\Omega^2/4\Delta$, and hence the loss drops by $\Omega^{-2}$. In the intermediate regime, the counter-rotation corresponding to the third order of nonlinearity slows the process down and hence reduces the loss rate reduction. Later the Rydberg population in the blockade radius reaches the maximum limit of one while the interaction keeps enhancing $\propto \Omega$  in the strong dressing regime leading to loss suppression scales by $\Omega^{-1}$. (b) The Poissonian probability of not having any BBR-induced depopulation $P_{BBR}(0)$ is plotted as a function of dressing strength. Going to resonance is important specially for large $N$ cat states to ensure the operation time finishes before the collective avalanche loss starts.   (c)  The outcome infidelity as a function of dressing strength. States with larger $m$ are less sensitive to the effects of higher-order nonlinearities. The cryogenic environment with T=77K is considered in calculations of (a,b).
}\label{Fig_IntToLoss}
\end{figure*}

For the weak dressing, where the third order of nonlinearity is negligible, the operation time $t_m=2\pi/m\chi$ would match perfectly with the numerical simulation used in Fig.~\ref{Fig_IntToLoss}. Going to strong dressing the operation time would be longer than $t_m$. This is because the third order of nonlinearity has the opposite sign with the second order, see Eq.~\ref{Eq_LightShift}, which makes the process a bit slower. However, the trend remains positive in terms of enhancing coherence.

  \begin{figure}
\centering    
        \subfloat{%
    \includegraphics[width=.49\textwidth]{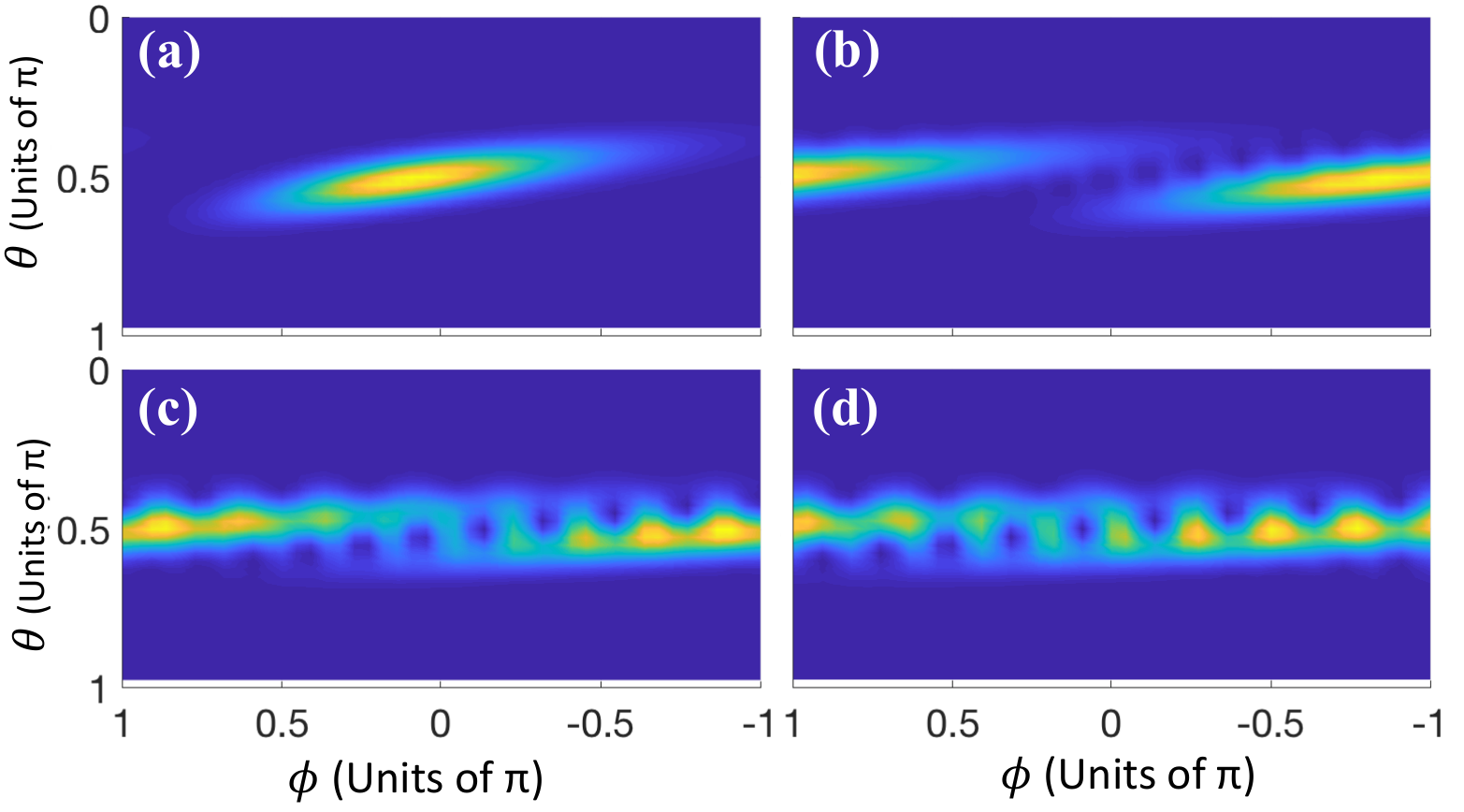}} \hfill         
\caption{The Physics of Multi-Component Cat Creation: The quadratic interaction term leads to varying rotation speeds around $J_z$ for different $|N_e\rangle$ elements, spanning the range $N_e^{max}-N_e^{min}=2\sqrt{N}$. This causes: (a) stretching of the initial CSS over the equator, (b-c) constructive and destructive interferences as the squeezed state's head and tail meet and pass through each other, and (d) formation of the first $m$-CSS superposition after a rotation difference of $\Delta \phi=4\pi$, allowing for interference to spread all over the equator.}\label{Fig_interfere}
\end{figure}

 Here we discuss the physics that determines the maximum number $m$ of CSS that could be formed in a superposition state. 
The population difference of the $N_e$ in the initial CSS is given by $N_e^{max}-N_e^{min}=2\sqrt{N}$. Hence the quadratic term of interaction would cause different rotation speeds around $J_z$ for different $\ket{N_e}$ elements of the initial CSS of Eq.~\ref{Eq_ICSS}.  
This would stretch the initial CSS over the equator. After the head and tail of the squeezed state meet and go through each other they form constructive and destructive interferences, which is shown as the superposition of CSSs, see Fig.~\ref{Fig_interfere}. 
Hence, the minimum required time to form the superposition is defined by the time that the head and tail of the squeezed state are stretched over $\Delta \phi=4\pi$  to spread the interference all over the Bloch sphere equator.
Considering the difference of speed under the quadratic term, the minimum dressing time that is required for the superpositions to appear would be $t_{\text{min}}=\frac{4\pi}{(N_e^{max}-N_e^{min})\Omega^4/16\Delta^3}=\frac{4\pi}{2\sqrt{N}\chi}$. 
The $\ket{m-\text{CSS}}$ would only be formed if its operation time $t_m=2\pi/m\chi$  occurs after the spread of interference all over the equator at $t_{\text{min}}$, see Fig.~\ref{Fig_interfere}.  Hence the maximum number of $m$-CSS superpositions that could be formed would be determined by the number of spins in the operation $m=\sqrt{N}$. One should note that this argument is derived at the weak dressing regime and going to strong dressing changes the operation times.  

\section{Advantage of strong dressing}
\label{Sec_Adv}

Cat states are extremely fragile with respect to decoherence, where a single decay leads to total destruction of entanglement. The decoherence is the bottleneck that limits cat states to tens of atoms in atomic systems. 
For constant ensemble size $N$, going to strong dressing with larger $\Omega/\Delta$ is favorable for enhancing the interaction-to-loss ratio.
To give a simple argument, lets consider a weak dressing regime, where the dominant term of interaction in the soft-core is $ \chi N_e^2= N_e^2\Omega^4/16\Delta^3$ while the loss rate from the Rydberg population $P_r=N_e(\frac{\Omega}{2\Delta})^2$ is given by $P_r\gamma_r $ with $\gamma_r$ being the loss rate from the Rydberg levels \cite{Bet09}. Considering the symmetry of states around the Bloch sphere's equator the $N_e$ could be replaced by $N/2$ in this scaling argument. 
Hence the interaction to loss ratio scales by $\frac{N\Omega^2}{4\Gamma \Delta}$. 
When transitioning to strong dressing, it becomes necessary to incorporate higher orders of expansion in Eq.~2. While the inclusion of odd terms may cause a deviation in the scaling from the specified interaction-to-loss ratio, the overall trend still favors moving towards strong dressing, as demonstrated below.

 Figure \ref{Fig_IntToLoss}a plots the Rydberg depopulation over the $\ket{m-\text{CSS}}$ generation as a function of dressing strength.   For simplicity, both the spontaneous emission and BBR-induced depopulation are considered as loss terms, which adds up to 4.8ms$^{-1}$ decoherence rate for the $|53P_{3/2},3/2\rangle$. Considering a 2D lattice with lattice constant $a=532$nm \cite{Wei11}, the laser detuning of $\Delta/2\pi=20$MHz accommodates  $N=48$ atoms well within the soft-core area $R_b/2$.  
At very weak dressing with quadratic nonlinearity, the interaction to loss ratio scales by $\Omega^2/4\Delta$, and hence the loss drops by $\Omega^{-2}$. In the intermediate regime, the counter-rotation effects of the third order of nonlinearity suppress the interaction and hence slow down the rate of loss reduction. at stronger dressing, the Rydberg population in the blockade radius reaches the maximum limit of one after which the loss rate remains constant. This is while the interaction keeps enhancing $\propto \Omega$  in the strong dressing regime, see Eq.~\ref{Eq_ResExp},  leading to loss suppression that is scaled by $\Omega^{-1}$.

Other than the spontaneous emission and BBR-induced depopulation of Rydberg states discussed above, experiments \cite{Zei16,Gol16} have observed a collective decoherence that is triggered by the BBR-induced depopulation of the Rydberg atoms. The BBR-induced depopulation to neighboring $n'P$ Rydberg states invokes a strong resonant dipolar interaction with the targeted $nS$ state resulting in an anomalous line broadening. This would further enhance the depopulation rate leading to a collective avalanche decoherence. Fig.~\ref{Fig_IntToLoss}b plots the Poissonian probability of not losing any Rydberg atom due to the BBR-induced depopulation $P_{BBR}(0)=\exp(-P_r\Gamma_{BBR} t_m)$ where the value of $\Gamma_{BBR}$ could be found in \cite{Bet09}. 
Reference \cite{Zei16} has observed that the avalanche decoherence only starts when $P_{BBR}(0)$ drops below 82\%. Fig.~\ref{Fig_IntToLoss}b shows the reduction of $P_{BBR}(0)$ by going towards the strong dressing. Going towards resonance would be vital for operations on large atom numbers in dense 3D lattices.

While the transition to a strong dressing regime is crucial for preventing decay over large-scale cat creation time, the mixed nonlinear terms lead to deviations from the targeted state. Figure \ref{Fig_IntToLoss}c compares the system state $\ket{\psi(t)}$ evolving under the exact Hamiltonian of Eq.~\ref{Eq_exact} with the targeted state $\ket{m-\text{CSS}}$ defined in Eq.~\ref{Eq_mCSS}. The fidelity is determined by optimizing the operation time and the parameters of the targeted state using the equation:
  \begin{equation}
\label{Eq_Fid}
F=\max_{\alpha_k,\phi_0,t} |\bra{\psi(t)} \frac{1}{\sqrt{m}}\sum\limits_{k=1}^{m} \text{e}^{i\alpha_k}   \ket{\theta=\pi/2;\phi=\frac{2\pi k}{m}+\phi_0}|^2.
\end{equation}
Fig.~\ref{Fig_IntToLoss}c exclusively considers the effects of mixed nonlinear terms without accounting for decoherence. The states with larger $m$ components are more robust against the presence of mixed nonlinearities. 
This has been quantified for the case of $\ket{2-\text{CSS}}$ and $\ket{6-\text{CSS}}$ in Fig.~\ref{Fig_IntToLoss}c but could not be calculated for larger $m$ due of the large dimension of optimization. However the Q function of $\ket{33-\text{CSS}}$ with N=1000 atoms generated under the resonant driving in Fig.~\ref{Fig_Resonance1000} shows a high fidelity outcome. Some examples of Hussimi Q-function at weak and strong dressing regimes are plotted in Fig.~\ref{Fig_QFWeakStrong} of App.~IV.

While transitioning to a stronger dressing deviates the outcome from the desired states, the sharp diving of the operation time and the loss rate that allows the generation of cat in the first place, outweigh the deviation from an exact $\ket{m-\text{CSS}}$ with large $m$.  
Consequently, depending on the applications, enhancing the coherence, or size of the entangled state may justify the reduction in fidelity.

\section{Going to resonance}
\label{Sec_Resonance}
  \begin{figure*}
\centering     
           \subfloat{%
    \includegraphics[width=1\textwidth]{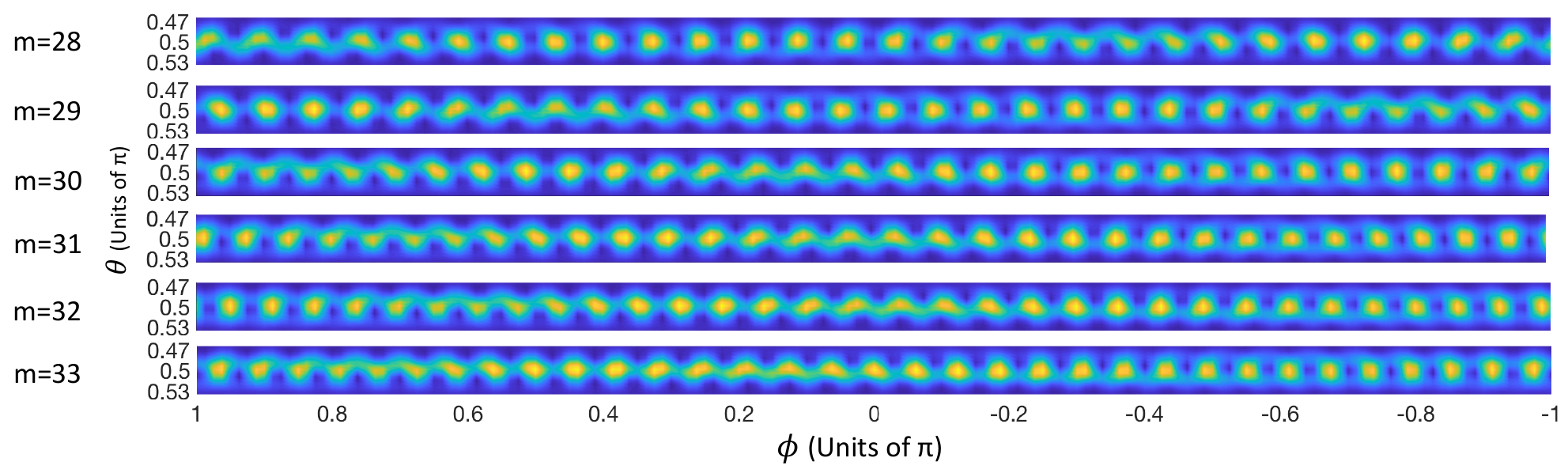}} \hfill          
\caption{Cat formation under resonance Rydberg driving Hamiltonian of Eq.~\ref{Eq_Res} applied on N=1000 atoms. }\label{Fig_Resonant}\label{Fig_Resonance1000}
\end{figure*}

{\bf  Resonant Rydberg driving Kerr Hamiltonian - }
As explained above approaching the resonance regime for the Rydberg exciting laser improves the interaction-to-loss ratio. This section discusses the cat generation under the resonance Rydberg excitation. Accommodating all the atoms inside the blockade radius the effective interaction would be given by
\begin{equation} 
\label{Eq_Res}
H_{res}=\sqrt{\hat{N}_e}\Omega
\end{equation} 
where $N_e$ is the number of atoms in $\ket{e}$ state.
Having a large number of atoms and initializing the CSS on the equator of the Bloch sphere, the average number of excited atoms $\bar{N}_e=N/2$ would be much larger than the deviation $\hat{N}_e-\bar{N}_e$ which is given by the radius of CSS i.e. $\sqrt{N}$. Hence in the regime of large atom number $N/2\gg\sqrt{N}$, the resonance Hamiltonian of Eq.~\ref{Eq_Res} could be expanded 
\begin{eqnarray}
\label{Eq_ResExp}
&&\hat{H}_{res}=\Omega\sqrt{\bar{N}_e}(1+\frac{\hat{N}_e-\bar{N}_e}{\bar{N}_e})^{{1/2}}\\ \nonumber
&&=\Omega\sqrt{\bar{N}_e}(1+\frac{\hat{N}_e-\bar{N}_e}{2\bar{N}_e}-\frac{(\hat{N}_e-\bar{N}_e)^2}{8\bar{N}_e^2}+\frac{(\hat{N}_e-\bar{N}_e)^3}{16\bar{N}_e^3} -...)\\ \nonumber
&&\approx \Omega\sqrt{\bar{N}_e} [\frac{5}{16}+\frac{15}{16}  \frac{ \hat{N}_e}{\bar{N}_e} - \frac{5}{16}(\frac{ \hat{N}_e}{\bar{N}_e})^2+\frac{1}{16}(\frac{ \hat{N}_e}{\bar{N}_e})^3-...]\\ \nonumber
&&\approx \Omega [\frac{5}{16}\sqrt{\frac{N}{2}}+\frac{15}{16} \sqrt{\frac{2}{N}} \hat{N}_e - \frac{5}{16}\sqrt{\frac{2^3}{N^3}}\hat{N}_e^2+\frac{1}{16}\sqrt{\frac{2^5}{N^5}}\hat{N}_e^3] \nonumber
\end{eqnarray}
 As an example, the formation of a few examples of $\ket{m-\text{CSS}}$ with $N=1000$ atoms under the resonant driving Hamiltonian of Eq.~\ref{Eq_Res} are plotted in Fig.~\ref{Fig_Resonant}. 
 The cat creation time obtained from the numerics are $\chi_2^{res}\times [t_{33},t_{27},t_{21},t_{14},t_{7},t_{4},t_3,t_2]=[0.236,0.289,0.373,0.563,1.132,2,2.655,4.061]$ which are normalized by the  dominant order of nonlinearity $\chi_{2}^{res}=\frac{5}{16}\sqrt{(\frac{2}{N})^3}\Omega$. The $\ket{33-CSS}$ state is formed 17 times faster than the conventional $\ket{2-CSS}$ cat state.

Given that the ensemble is located within the blockade radius, the Rydberg population is fixed at 1 in the resonant driving model. Consequently, the loss can be calculated as the product of the operation time and the Rydberg loss rate. 
For instance, let's consider the resonant driving of $N=1000$ $^{87}$Rb atoms to the $\ket{80S_{1/2}}$ state with a Rabi frequency of $\Omega/2\pi$=70MHz. In the cryogenic environment at $T=77$K, the Rydberg decay rate is $\Gamma_r=2400$ s$^{-1}$. Numerically, the operation times for creating the $\ket{33-CSS}$ and $\ket{2-CSS}$ states are found to be $t_{33}=0.236/\chi_{2}^{res}$ and $t_{2}=4/\chi_{2}^{res}$, respectively. The corresponding Poissonian probabilities of not losing any atom over the operation times $t_{33}$ and $t_{2}$ are 98\% and 66\%, respectively.
In a scaling argument, as the principal quantum number increases, the Rydberg decay rate is suppressed as $\Gamma_r \propto n^{-3}$, while the interaction is enhanced as $C_6\propto n^{11}$, allowing for stronger laser driving and faster operation for a constant blockade radius. Consequently, the atom loss over cat creation scales inversely with the principal number as $n^{-14}$ for a constant atom number $N$.

{\bf A remark on Adiabaticity --}
Considering the off resonant Rydberg dressing as explained in Sec.~\ref{Sec_Dress}, the laser couples two  states $\ket{\psi_0}$ and  $\ket{\psi_1}$ with collective Rabi frequency  $\sqrt{N_{e}}\Omega_{r}$ and detuning $\Delta$. The time evolution of dressed eigen states $\ket{\tilde{\psi}_{\pm}}$  would be given by 
\begin{equation}
\label{Eq_Ham}
i\frac{\partial}{\partial t} \left(\begin{array}{cc} \ket{\tilde{\psi}_{-}}   \\ \ket{\tilde{\psi}_{+}} \end{array}\right)=\left(\begin{array}{cc}\mathcal E_{-} & -i\dot{\theta}/2 \\ i\dot{\theta}/2 & \mathcal E_{+}\end{array}\right) \left(\begin{array}{cc} \ket{\tilde{\psi}_{-}}  \\ \ket{\tilde{\psi}_{+}} \end{array}\right).
\end{equation}
where $\mathcal E_{\pm}=\frac{\Delta}{2}(1 \pm \sqrt{1+\frac{N_{e}\Omega^{2}}{\Delta^2}})$ are the energies of ground and excited dressed states and $\dot{\theta}=\frac{\sqrt{N_e}\Omega \dot{\Delta} -\sqrt{N_e} \Delta \dot{\Omega} }{N_e \Omega^2+\Delta^2}$.
In the case of non-zero detuning, it is important to keep the off-diagonal terms small to minimize the population scattering to the other eigenstate which is quantified by $\dot{\theta}^2/E_{+}^2$. The scattered population would remain in the Rydberg state after switching of the laser which results in the distortion of cat states.
On the other hand, the off-diagonal terms would not appear when the laser gets in resonance with the Rydberg level. This would bring the advantage of fast switching of the laser in resonant driving. 
 
\section{conclusion}
In the thriving field of Rydberg technology \cite{Mor21,Kha20,khaz2020rydberg,Kha19,KhaRev21,Kha22,Khaz21Fermi,Kha23SubNano,Kha22Stab}, the Rydberg dressing plays an important role in the implementation of quantum matters and making large scale entanglement.
This article delves into the generation of multi-component cat states through both strong Rydberg dressing and resonant Rydberg driving, departing from previous research that primarily focused on the weak dressing regime to isolate the second order of nonlinearity. By approaching resonance, the interaction-to-loss ratio is enhanced, enabling successful operation termination before the onset of BBR-induced collective avalanche decoherences. The findings here demonstrate that resonant Rydberg driving can effectively isolate quadratic nonlinearity with a large number of atoms accommodated within the blockade volume.

Moreover, this paper introduces a novel method to create multi-component cat states in significantly shorter timescales compared to traditional two-component cat states. These states exhibit reduced sensitivity to mixed nonlinearity orders, making them an attractive option for generating large-scale entangled states crucial for applications in metrology and quantum error correction. Lastly, App.~I  presents a perturbative analytic formula for strong dressing interactions, suggesting the optimization of lattice geometry to resonate and enhance specific orders of nonlinearity. 

{\bf Acknowledgement:} The author acknowledges motivating discussions with Nathan Schine.

\section*{Appendix I: Many-body interaction at strong dressing }
\label{sec:PerturbDressing}

To investigate the strong dressing regime in mean field formalism, one needs analytic formula for the interaction profile.  
The profile of weak dressing interaction has been formulated in a perturbative approach that covers up to the two-body interaction \cite{Hen10}.
 Going to stronger dressing, the effects of higher-order terms would be magnified. This section looks into the analytical profile of the interaction up to the third order of nonlinearity.

Considering $N$ atoms in an optical lattice, for each pair of atoms excited to the Rydberg level $|r\rangle$ and separated by ${\bf x}_{ij}={\bf x}_i - {\bf x}_j$, where ${\bf x}_i$ is the position of the $i$th atom, the binary interaction is
$V_{ij}=C_6/x_{ij}^6$. Here the quantization axis is considered perpendicular to the lattice plane to preserve the isotropy of interaction. 
The dressing potential $U({\bf x}_1..{\bf x}_{N_e})$ of the state $|\psi({\bf x}_1..{\bf x}_{N_e}) \rangle$ with $N_e$ atoms in $\ket{e}$ being dressed with Rydberg level $\ket{r}$, is calculated  under the condition of $(\frac{\Omega}{2 \Delta})^2 \ll 1$  by applying  perturbation theory
\begin{equation}
\frac{U}{\Delta}|\psi\rangle=
({\sum\limits _{i=1}^{N} \hat \sigma_{rr}^{i}+\frac{1}{\Delta} \sum\limits _{i<j} V_{ij} \hat{\sigma}_{rr}^{i}\hat{\sigma}_{rr}^{j}}) +{{\frac{\Omega}{2\Delta}}} {\sum\limits _{i=1}^{N}(\hat \sigma_{re}^{i}+\hat \sigma_{er}^{i}}) \, |G\rangle,
\end{equation}
where the first and second parentheses separate the unperturbed and perturbed parts of Hamiltonians.

The contribution from the second and fourth-order perturbations are calculated as \cite{Hen10}
\begin{equation}
U^{(2)}=-\frac{\Omega^2}{4\Delta}N; \quad  U^{(4)}=
 \frac{\Omega^4}{16\Delta^3}\sum\limits _{i<j}[\frac{1}{(1+\frac{V_{ij}}{2\Delta})}-1]
\label{Eq_4thOrder}
\end{equation}
where the former is the sum of the single atom light-shift and the latter encounters the two-body interactions among atoms.  
\begin{figure*}
\centering        
        \subfloat{%
    \includegraphics[width=1\textwidth]{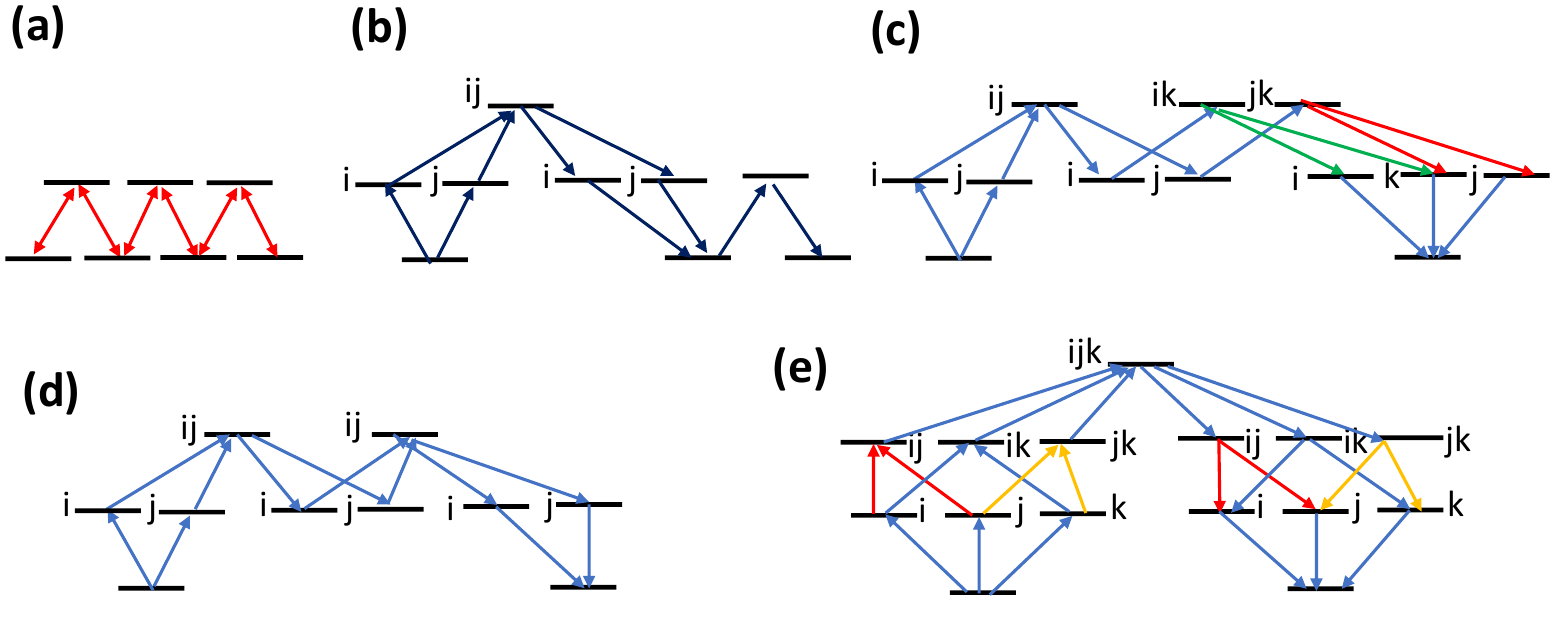}} \hfill
\caption{Perturbation paths used to calculate the $U^{(6)}$ in Eq.~\ref{Eq_6thOrder}. Indices $i,j,k$ label distinguished atoms excited to the Rydberg state. }\label{Fig_Perturbation}
\end{figure*} 
In the limit of strong interaction ($V\rightarrow \infty$) these terms reproduce the first two terms of collective light-shift expansion of Eq.~\ref{Eq_LightShift}.

 Going to a stronger dressing regime with a non-negligible third order of nonlinearity $(N_e\frac{\Omega^2}{4\Delta^2})^3$ the sixth-order of perturbation must be taken into account.  Here different configurations of the six photon transitions in the perturbative ladder are plotted in Fig. \ref{Fig_Perturbation}, which includes the two-body as well as three-body interactions. The sixth correction of the interaction profile will be given by
 \begin{widetext}
\begin{eqnarray} 
  \label{Eq_6thOrder}
 U^{(6)}=&&\frac{\Omega^6}{2^6\Delta^5}[-2N^3+(\sum_{k=1}^N(1))\sum_{i\neq j}\frac{4}{2-V_{ij}/\Delta}+\sum_{i\neq j} \frac{2}{2-V_{ij}/\Delta} (\sum_{k\neq i}\frac{2}{2-V_{ik}/\Delta}+\sum_{k\neq j}\frac{2}{2-V_{jk}/\Delta})\times2\\ \nonumber
  &&-\sum_{i\neq j} \frac{2}{2-V_{ij}/\Delta} (\sum_{i}\frac{1}{2-V_{ij}/\Delta}+\sum_{j}\frac{1}{2-V_{ij}/\Delta})-\sum_{(ijk)\neq} (\frac{2}{2-V_{ij}/\Delta}+\frac{2}{2-V_{ik}/\Delta}+\frac{2}{2-V_{jk}/\Delta})^2\times\\ \nonumber
 && \frac{1}{3-V_{ij}/\Delta-V_{ik}/\Delta-V_{jk}/\Delta}]
 \end{eqnarray}
 \end{widetext}
Comparing the terms in Eq.~\ref{Eq_4thOrder} and \ref{Eq_6thOrder}, one can see that the ratio of $\Delta/V$ in a lattice with a given geometry could be used as a knob to control the relative strength of different nonlinear terms. For example, expanding this trend to the higher order one can see that in a square/triangle lattice, the fourth/sixth order of non-linearity would get enhanced around $\Omega=\Delta$. 
This could be instrumental in isolating the fourth order of non-linearity with potential applications in cat state error correction codes \cite{Pur20}.

   \section*{App. II: Numerical calculation of the interaction profile}

      The level scheme depicted in Fig.~\ref{Fig_Scheme}a  pertains to three-level Rubidium atoms undergoing off-resonance excitation to the highly excited Rydberg state $|r\rangle$. The laser driving Hamiltonian for the $i$-th atom is given by
$H_i=\frac{\Omega}{2} (\hat{\sigma}^i_{re}+\hat{\sigma}^i_{er})-\Delta \hat{\sigma}^i_{rr}$, 
where $\sigma_{\alpha,\beta}=|\alpha \rangle\langle \beta|$,  $\Omega$ and $\Delta$ represent the Rabi frequency and detuning of the transition, respectively. The van der Waals interaction between Rydberg atoms, denoted as $V_{ij}=C_6/r_{ij}^6 \sigma_{rr}^i \sigma_{rr}^j$, is a function of the interatomic distance $r_{ij}$. The total multi-atom dressing Hamiltonian is expressed as $H_d=\sum_i\hat{H}_i+\sum_{i<j}V_{ij}$.
The dynamics of the system under the Rydberg dressing interaction is governed by the master equation, which can be represented as
\begin{eqnarray}\label{Eq_Heisenberg}
\partial_t \hat{\rho}=-\text{i}[H_d,\hat{\rho}] + \sum_i\mathcal{L}_i(\hat{\rho}) \\ \nonumber
\end{eqnarray}
where the Liouvillian operator $\mathcal{L}_i (\rho)=c\rho_ic^{\dagger}-1/2 (c^{\dagger}c \rho_i+\rho_ic^{\dagger}c )$ in Lindblad form describes the single-particle dissipation affecting the internal state dynamics, with $c=\sqrt{\gamma_r }|e\rangle \langle r|$ governing the spontaneous emission from the Rydberg state.

Upon considering the steady state $\rho$ from Eq.~\ref{Eq_Heisenberg}, the interaction can be calculated as
\begin{equation}
U= Tr[\rho H_d]
\end{equation}
Figure \ref{Fig_Scheme}b illustrates the dressing interaction profile for two atoms (solid lines) and three atoms (dashed lines) arranged in an equilateral triangle with sides of length $r$ at various dressing strengths. To present the effective interaction, the background interaction-independent light-shift $U (r=\infty)$, which only generates a constant phase, is subtracted. The blockade radius is defined as $R_b=(C_6/\Delta)^{1/6}$ in the weak dressing regime and as $R_b=(C_6/\Omega)^{1/6}$ in the strong dressing regime. It is evident from figure \ref{Fig_Scheme}b that in both weak and strong dressing, as long as the atoms are within the soft-core with a radius of $R_b/2$, the interaction becomes independent of atomic distance and is solely defined by the collective light shift, characterized by the number of atoms and laser driving parameters.

 \begin{figure*}
\centering        
        \subfloat{%
    \includegraphics[width=1\textwidth]{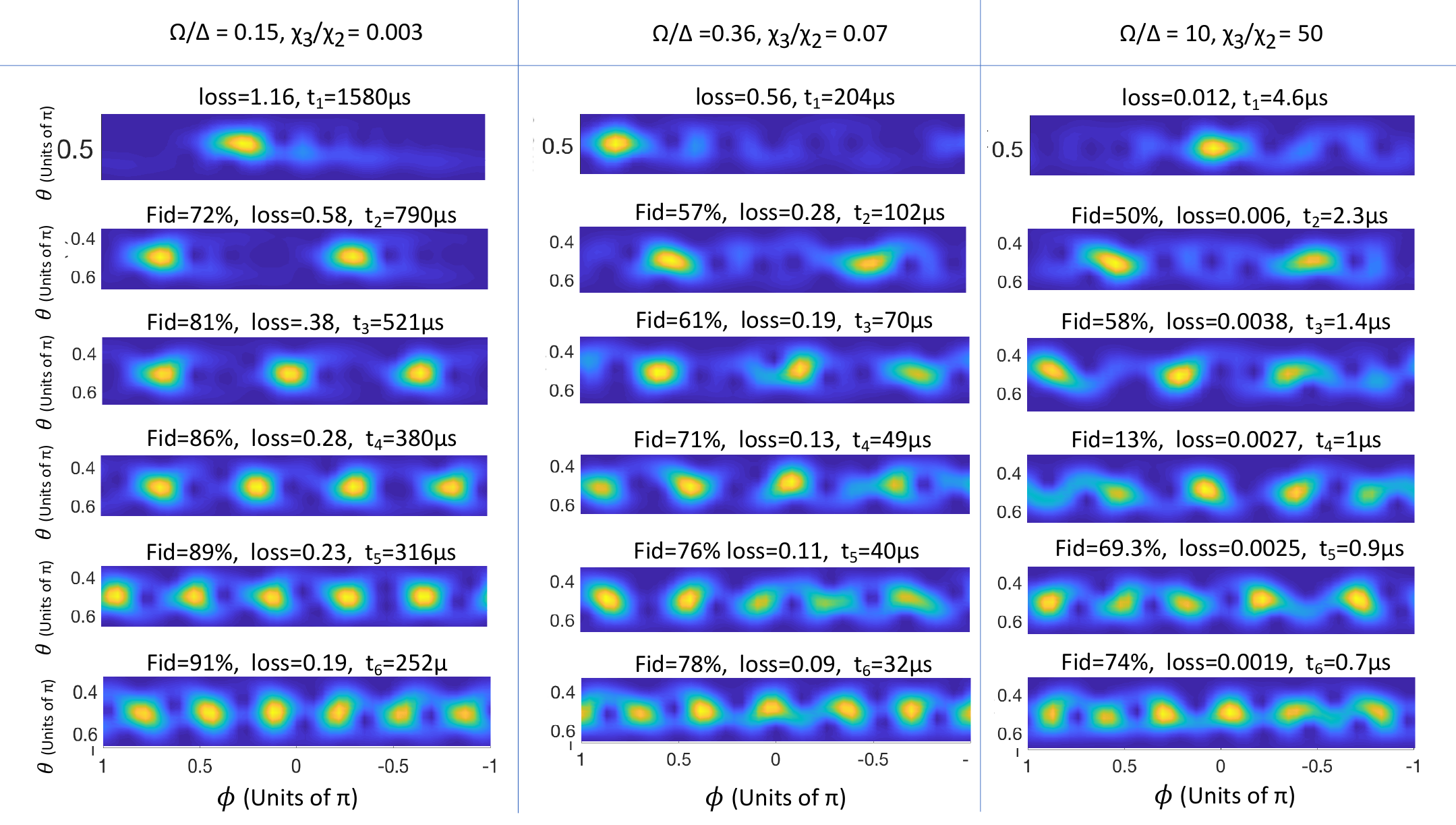}} \hfill
\caption{  Transition from a weak to a strong dressing regime. The revival of a single CSS at time $t_1$ indicates the coherent superposition of $|m\text{-CSS}\rangle$ states in both the weak and strong dressing regimes. The last column contains an ascending contribution of all nonlinear terms in Eq.~\ref{Eq_LightShift}. }\label{Fig_QFWeakStrong}
\end{figure*}

\section*{App. III: Assessing Coherence through Evolution Analysis}

 While the Wigner representation effectively visualizes the coherence of superposition in photonic systems through the presence of fringes in phase space, this method cannot be directly applied to coherent spin states (CSS) in atoms.
Instead, the Hussimi Q-function offers a convenient means of visualizing quantum states by projecting them onto the coherent spin states $\ket{\theta,\phi}$, where the parameters span the Bloch sphere. 
However, it is important to note that the Hussimi Q-function alone does not inherently distinguish a cat state, representing a coherent superposition state, from a mixed state. To assess coherence, one can examine the retrieval of a single CSS under the dressing interaction at $t_1$, as discussed below.

 Let's analyze the evolution of the CSS under the second-order nonlinearity $\chi \hat{N}_e^2$.
  As previously described, the initial CSS $\ket{\eta}$ evolves into $(e^{-i\pi/4}\ket{\eta}+e^{i\pi/4}\ket{-\eta})/\sqrt{2}$ after $t_2$ duration of dressing (Recal $t_m=\frac{2\pi}{m\chi}$). By observing the evolution of the density matrix in the {$\ket{\eta}$, $\ket{-\eta}$} basis, the initial state $\rho(0)=\begin{pmatrix} 1 & 0\\ 0 & 0 \end{pmatrix}$ evolves into $\rho(t_2)=1/2\begin{pmatrix} 1 & i\\ -i & 1 \end{pmatrix}$.
 Subsequently, the same map transfers the state to $\rho(t_1)=\begin{pmatrix} 0 & 0\\ 0 & 1 \end{pmatrix}$ after another $t_2$  dressing time line (Note $t_1=2t_2$).  Conversely, if the state at $t_2$ was a mixed state $\rho_{\text{mixed}}(t_2)=1/2\begin{pmatrix} 1 & 0\\ 0 & 1 \end{pmatrix}$, it would evolve into another statistical mixture of CSSs $\rho_{\text{mixed}}(t_1)=1/2\begin{pmatrix} 1 & 0\\ 0 & 1 \end{pmatrix}$ after the same duration of dressing time.
 
In conclusion, while the Hussimi Q-function of the cat state $\rho(t_2)$ and a mixed state $\rho_{\text{mixed}}(t_2)$ may not be distinguishable, the revival of a CSS at $t_1$  provides a clear signature of coherence at the earlier time, see Fig.~\ref{Fig_QFWeakStrong}. The coherent cat state would evolve into a single CSS, whereas a statistical mixture of CSSs at $t_2$ would evolve into another mixture of separate peaks at $t_1$.

\section*{Appendix IV: Depiction of cat states' Q-function in weak and strong dressing }

Some examples of Hussimi Q-function of simulated cat states under the exact Hamiltonian of Eq.~\ref{Eq_exact}  at weak and strong dressing regimes are plotted in Fig.~\ref{Fig_QFWeakStrong}. 
Here a 2D lattice with lattice constant $a=532$nm \cite{Wei11} is considered. Dressing atoms with $|53P_{3/2},3/2\rangle$ Rydberg state with laser detuning of $\Delta/2\pi=0.02$GHz accommodates  $N=48$ atoms well within the soft-core area $R_b/2$.    In Fig.~\ref{Fig_QFWeakStrong}, the first column corresponds to cases where the second order of nonlinearity $\chi_2$ is isolated, in the second column partial involvement of the third order $\chi_3/\chi_2=0.06$ is considered, and in the third column, the extreme case is applied where all orders of nonlinearity are involved in ascending order. 
The cat formation process could be observed in the case that all orders of nonlinearity exist, however with reduced fidelity.

The applied $\ket{m-\text{CSS}}$ that are used for defining the fidelity in Fig.~\ref{Fig_QFWeakStrong} and Fig.~\ref{Fig_IntToLoss}c of the main paper are as following:
    \begin{widetext}

    $\ket{2-CSS}=[e^{-i\pi/4}\ket{\eta}+e^{i\pi/4}\ket{e^{i\pi}\eta}]/\sqrt{2}$

      $\ket{3-CSS}=[e^{-i\pi/3}\ket{\eta}+e^{-i5\pi/3}\ket{e^{i2\pi/3}\eta}+e^{-i5\pi/3}\ket{e^{i4\pi/3}\eta}]/\sqrt{3}$
      
      $\ket{3-CSS}=[e^{-i2\pi/3}\ket{\eta}+e^{-i\pi/3}\ket{e^{i2\pi/3}\eta}+e^{-i\pi/3}\ket{e^{i4\pi/3}\eta}]/\sqrt{3}$
      
      $\ket{3-CSS}=[e^{i2\pi/3}\ket{\eta}+e^{-i2\pi/3}\ket{e^{i2\pi/3}\eta}+e^{-i2\pi/3}\ket{e^{i4\pi/3}\eta}]/\sqrt{3}$
      
      $\ket{3-CSS}=[e^{i\pi/3}\ket{\eta}+e^{-i\pi}\ket{e^{i2\pi/3}\eta}+e^{-i\pi}\ket{e^{i4\pi/3}\eta}]/\sqrt{3}$

      $\ket{4-CSS}=[e^{-i\pi/4}\ket{\eta}+\ket{e^{i2\pi/4}\eta}+e^{-i\pi/4}\ket{e^{i4\pi/4}\eta}+e^{-i\pi}\ket{e^{i6\pi/4}\eta}]/\sqrt{4}$
      
      $\ket{5-CSS}=[e^{-i4\pi/5}\ket{\eta}+e^{-i8\pi/5}\ket{e^{i2\pi/5}\eta}+e^{-i4\pi/5}\ket{e^{i4\pi/5}\eta}+e^{-i2\pi/5}\ket{e^{i6\pi/5}\eta}+e^{-i2\pi/5}\ket{e^{i8\pi/5}\eta}]/\sqrt{5}$
      
      $\ket{5-CSS}=[e^{-i8\pi/5}\ket{\eta}+e^{-i2\pi/5}\ket{e^{i2\pi/5}\eta}+e^{-i8\pi/5}\ket{e^{i4\pi/5}\eta}+e^{-i6\pi/5}\ket{e^{i6\pi/5}\eta}+e^{-i6\pi/5}\ket{e^{i8\pi/5}\eta}]/\sqrt{5}$
      
      $\ket{6-CSS}=[\ket{\eta}+e^{-i11\pi/6}\ket{e^{i2\pi/6}\eta}+\ket{e^{i4\pi/6}\eta}+e^{-i3\pi/6}\ket{e^{i6\pi/6}\eta}+e^{-i8\pi/6}\ket{e^{i8\pi/6}\eta}+e^{-i3\pi/6}\ket{e^{i10\pi/6}\eta}]/\sqrt{6}$
   \end{widetext}


\begin{thebibliography}{100}
\bibitem{abnormalOrdering} There are specific cases where $\ket{m-\text{CSS}}$ appears at times that do not follow the orders mentioned above. For example in the following DOI 10.6084/m9.figshare.24851724 in video "N37D0.3w.mov", see frame 61. Other examples are in  "D0.1w0.2.mov" at frame 56 and in "D0.01w0.1.mov" in frames 25 35, 49.
\end{thebibliography}
\end{document}